\documentclass[10pt,conference]{IEEEtran}
\IEEEoverridecommandlockouts

\usepackage[normalem]{ulem}
\usepackage{algorithm}
\usepackage{braket}
\usepackage[noend]{algpseudocode}
\usepackage{amssymb}
\usepackage[dvipsnames]{xcolor}
\usepackage{float}
\usepackage{placeins} % for placement of graphs
\usepackage{booktabs}
\usepackage{multirow}
\usepackage{graphicx}
\usepackage{fancyhdr}
\usepackage{tabularx}
\usepackage{booktabs}
\usepackage{caption}
\usepackage{dblfloatfix}
\usepackage{pbox}
\usepackage{version}

\usepackage[top=1in, bottom=1in, left=1in, right=1in]{geometry}
\usepackage{graphicx,amsmath,amsthm,cite,nicefrac,ifthen,amsfonts, amssymb}
\usepackage{subcaption}
\usepackage{wrapfig}

\usepackage{subcaption}
\usepackage{enumitem}
\setlist{leftmargin=4mm}

\usepackage{soul,xcolor}
\setstcolor{red}
\newcommand{\getCurrentSectionNumber}{%
  \ifnum\c@section=0 %
  \thechapter
  \else
  \ifnum\c@subsection=0 %
  \thesection
  \else
  \ifnum\c@subsubsection=0 %
  \thesubsection
  \else
  \thesubsubsection
  \fi
  \fi
  \fi
}

\theoremstyle{definition}

\theoremstyle{definition}

\theoremstyle{definition}

\makeatletter
\def\tagform@#1{\maketag@@@{\bfseries(\ignorespaces#1\unskip\@@italiccorr)}}
\renewcommand{\eqref}[1]{\textup{{\normalfont(\ref{#1}}\normalfont)}}
\makeatother
\begin{document}

\pagestyle{fancy}
\fancyhf{}  
\fancyhead[R]{\thepage} % Page number on the top-right
\renewcommand{\headrulewidth}{0pt}  
\renewcommand{\footrulewidth}{0pt} 
\thispagestyle{fancy}

\title{Optimizing Orbital Parameters of Satellites for a Global Quantum Network
%\\\thanks{This research was supported in part by the NSF grants CNS-1955744 and 1934846, NSF-ERC Center for Quantum Networks grant EEC-1941583, and the MURI ARO Grant W911NF2110325}}
}

%\author{\IEEEauthorblockN{Anonymous Authors}} 
\author{\IEEEauthorblockN{Athul Ashok\IEEEauthorrefmark{1}\IEEEauthorrefmark{2}, Owen DePoint\IEEEauthorrefmark{1}\IEEEauthorrefmark{2}, Jackson MacDonald\IEEEauthorrefmark{1}, Albert Williams\IEEEauthorrefmark{1}, and Don Towsley\IEEEauthorrefmark{1}}\IEEEauthorblockA{\IEEEauthorrefmark{1}University of Massachusetts, Amherst, MA, USA \\Email: \IEEEauthorrefmark{1}\{aashok@, odepoint@, jrmacdonald@, abwillia@cs., towsley@cs.\}umass.edu}\IEEEauthorrefmark{2}Contributed equally to this work}
%\IEEEauthorrefmark{1}

\maketitle

 \begin{abstract}
 Due to fundamental limitations on terrestrial quantum links, satellites have received considerable attention for their potential as entanglement generation sources in a global quantum internet. In this work, we focus on the problem of designing a constellation of satellites for such a quantum network. We find satellite inclination angles and satellite cluster allocations to achieve maximal entanglement generation rates to fixed sets of globally distributed ground stations. Exploring two black-box optimization frameworks: a Bayesian Optimization (BO) approach and a Genetic Algorithm (GA) approach, we find comparable results, indicating their effectiveness for this optimization task. While GA and BO often perform remarkably similar, BO often converges more efficiently, while later growth noted in GAs is indicative of less susceptibility towards local maxima. In either case, they offer substantial improvements over naive approaches that maximize coverage with respect to ground station placement.
 \end{abstract}

%\section{Abstract}

\section{Introduction}\label{sec:intro}
Satellite-based quantum networks enable long-distance quantum communication by leveraging satellites and ground stations to distribute quantum entanglement over long distances. Unlike fiber optic cables or terrestrial free-space channels, satellite links offer significantly lower transmission losses over distances greater than 100 kilometers, making them favorable for large-scale quantum communication \cite{10713083}.
Optimizing orbital parameters, such as the inclination angle \cite[p.~158]{curtis2015orbital} of the orbit above the equator and the number of satellites within a constellation, is crucial for maximizing the coverage of the network to ground stations. Maximal ground station coverage ensures consistency and reliability within the network when implemented on a large scale.

In the dual downlink architecture, satellites generate entangled qubit pairs and transmit them simultaneously to two ground stations via free-space links. This setup requires that both ground stations be simultaneously within the same satellite's line of sight, limiting connections to geographically proximate stations. Although this architecture imposes constraints on network connectivity and necessitates efficient scheduling to optimize entanglement distribution rates, it eliminates the need for quantum memories onboard satellites \cite{williams2024scalableschedulingpoliciesquantum}. We simulate a Spontaneous Parametric Down-Conversion (SPDC) entanglement generation source because it is an experimentally validated and compact method for quantum satellite missions. SPDC has been successfully implemented in orbit, i.e. on the SpooQy-1 CubeSat \cite{Sivasankaran_2022}.

Williams et al. \cite{williams2024scalableschedulingpoliciesquantum} and Karavias et al. \cite{Karavias:24} both examine the problem of optimizing the allocation of satellites to ground stations in a quantum network, the former maximizes the performance of an existing constellation, and the latter attempting to minimize the number of satellites needed in a constellation. Our work instead optimizes the placement of satellites by altering orbital elements of satellites, in this case, the satellite inclinations, within a constellation to maximize performance. We show that black-box optimization tools can be used to define satellite constellations that outperform naive, regular approaches which attempt to maximize coverage without respect to ground station placement.

We select two well-established lightweight black-box optimization methods: Bayesian Optimization (BO) and Genetic Algorithms (GA), each implemented with hyperparameters aligned with prior work on similarly high-dimensional, computationally expensive problems. Specifically, we use a BO implementation that employs Gaussian process surrogate modeling and standard acquisition functions (expected improvement and lower confidence bound) \cite{ frazier2018tutorialbayesianoptimization, 9696089} and a GA implementation that makes use of hyperparameters such as exponentially decaying mutation rate and multi-parent crossover \cite{10.5555/93126, 10.1007/3-540-58484-6_252}. Both methods were carefully tuned to effectively navigate the complex search space, while also efficiently addressing computational constraints when interacting with a quantum network simulation.

In the classical navigational satellite constellation setting, Jiang et al. \cite{JIANG2024} demonstrates BO to be a highly effective tool for optimizing orbital parameters, with computational advantages over traditional methods, such as GA. However, our quantum networking setting introduces additional structural constraints stemming from quantum hardware cost, dual downlink architecture, and qubit fragility during transmission.

Our results demonstrate that both BO and GA are reliable and scalable frameworks for optimizing satellite constellations for quantum networks. Across all tested scenarios, the optimized constellations consistently outperform regular designs by significant margins, achieving upwards of $2.09 \times 10^{6}$ EPR pairs per second under a population-based ground station placement. In contrast, naive strategies, including uniformly equispaced constellations, consistently performed 28-57\% lower. Our results demonstrate that known optimization algorithms can serve as an effective methodology for constellation design in quantum satellite networks. 

In Section \ref{sec:preliminaries} we begin by addressing preliminary physical problem constraints and constellation design choices. Section \ref{sec:opt-preliminaries} then details the general optimization problem that we attempt to solve, while Section \ref{sec:opt-methodology} defines and connects our optimization frameworks to the given problem. Section \ref{sec:assessment} details how we assess the effectiveness of these frameworks, and Section \ref{sec:results} analyzes the resulting numerical performance in relation to baselines and each other. Finally, Section \ref{sec:conclusion} concludes the paper with a summary and direction for future work.

\section{Preliminaries}\label{sec:preliminaries}
Our objective is to compare the efficacy of BO and GA against more naive approaches when designing a satellite constellation. We specifically focus on optimizing both inclination angles and number of satellites assigned to each inclination to maximize the average entanglement distribution rate per second. We simulate a 100 satellite constellation with orbits at an altitude of 550 kilometers and 100 ground stations, either placed over the most 100 populous cities (see Figure ~\ref{fig:sat-gs-info}(a)) or randomly on land (see Figure ~\ref{fig:sat-gs-info}(b)). We use the expected number of EPR pairs distributed to a given set of ground stations per second over the course of a day, with samples taken every 30 seconds, as the metric for evaluating a constellation.

We evaluate constellations with 1, 2, 3, and 5 separate orbits with variable inclination angles and variable numbers of satellites assigned to each orbit. We assume each satellite has one pair of transmitters and each ground station has one receiver, and we use the \texttt{GREEDY\_BACKOFF} algorithm described in \cite{williams2024scalableschedulingpoliciesquantum} to choose which pairs of ground stations communicate with which satellites.

In this work, we consider a dual downlink architecture for entanglement distribution (see Section \ref{sec:intro}). We use the SPDC source and atmospheric channel model defined in \cite{panigrahy2022optimalentanglementdistribution}, with the noise and loss model based on two pair photon emissions and background photon flux described in \cite{williams2024scalableschedulingpoliciesquantum}. As in \cite{williams2024scalableschedulingpoliciesquantum}, we require that satellites be at a minimum elevation angle of $20^\circ$ above the horizon of a ground station to be visible.

\section{Optimization Preliminaries}\label{sec:opt-preliminaries}

\begin{figure*}[htbp]
  \centering  \includegraphics[width=0.99\linewidth]{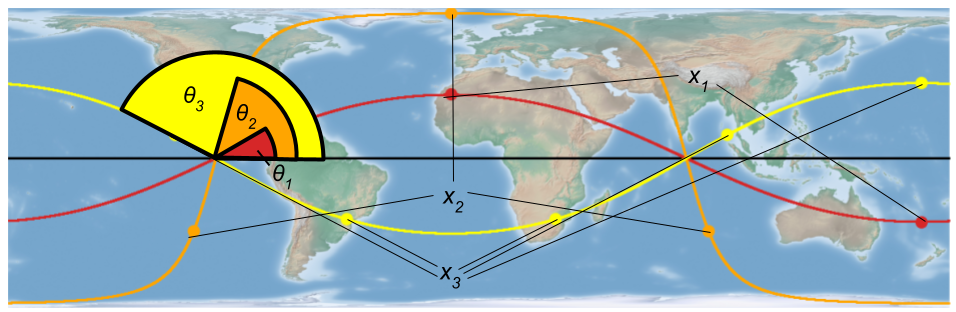}
  \captionsetup{justification=centering,width=0.99\linewidth}%
  \caption{%
    Example of a constellation with the hyperparameter $D=3$ and the following parameters: $\theta_1=25^\circ, \theta_2=75^\circ, \theta_3=150^\circ, x_1=2, x_2=3, x_3=5$
  }
  \label{fig:map_parameterization}
\end{figure*}

Before describing our BO and GA optimization frameworks, we outline the general parameterization and transformations that enable effective search over the space of possible satellite orbits. We focus on handling both continuous inclination angles and the constrained allocation of satellites into multiple orbits. We constrain the number of separate orbits that can be explored, requiring at each iteration all $N =100$ satellites be placed within one of our $D \in \{1, 2, 3, 5\}$ specified orbits.

\subsection{Parametrization of satellite constellations}
We vary our satellite constellations according to the following parameters:
\label{sec:parametrization}
\begin{itemize}
    \item \textbf{Inclination Angle:} Each orbit has an inclination $\theta_i \in [0^\circ, 180^\circ]$, where an inclination angle $\theta$ yields the same orbit as that of $180^\circ - \theta$, except reflected over the equator.
    \item \textbf{Satellite Allocation:} We distribute $N$ satellites across $D$ orbits, while ensuring our constraints are maintained. Namely, the assignment of satellites to orbits must respect the overall satellite count and must be constrained to a state space over $\mathbb{Z}^D$. To do this, let $\mathbf{x} =  [x_1, x_2, \dots, x_D]$ where $x_i \in \mathbb{R}$ represents the fraction of satellites allocated to the $i$-th orbit. We require
    $x_i \geq 0 \text{ and } \sum_{i=1}^{D} x_i = 1.$
\end{itemize}

Taking any value $x_i \in [0, 1]$, we map these to proportional amounts of satellites:
$
\text{s}_i = x_i \cdot N,
$
which are then mapped to integers (see Section C). If necessary, additional constraints (e.g., a minimum number of satellites per orbit) can be enforced by adjusting or penalizing invalid allocations. See Figure \ref{fig:map_parameterization} for an example visualization of this parameterization.

\subsection{The Additive Log-Ratio (ALR) Transformation}\label{sec:alr}

Enforcing the constraints $\forall x_i \geq 0$ and $\Sigma x_i = 1$ is not straightforward, as the number of satellites allocated to each orbit is specified independently (so we cannot optimize directly over the $D$-dimensional simplex). Instead, we proceed by inverting the Additive Log-Ratio (ALR) transformation. The ALR transformation maps the $D$-dimension simplex into an unconstrained $(D-1)$-dimensional real space, using one of the $D$ variables as a \textit{reference},
\[
y_i = \log_e\left(\frac{x_i}{x_D}\right) \quad \text{for } i = 1,\ldots,D-1,
\]
where $x_D$ plays the role of the reference component. We can, given a vector $\mathbf{y} \in \mathbb{R}^{D-1}$, retrieve our desired fractions $\mathbf{x}$ via the inverse transformation:
\[
x_i = \frac{e^{y_i}}{1 + \sum_{j=1}^{D-1}e^{y_j}} \quad \text{for } i = 1,\ldots,D-1,
\]
\[
x_D = \frac{1}{1 + \sum_{j=1}^{D-1}e^{y_j}}.
\]

This guarantees that any real vector $\mathbf{y}$ corresponds to a unique, valid point $\mathbf{x}$ in the simplex. We then proceed with standard unconstrained optimization methods, mapping these fractions as follows.

\subsection{From fractions to integer allocations}

Once we have the fractions $\mathbf{x} = (x_1,\ldots,x_D)$, we compute the integer satellite counts:
\[
\textit{sats\_per\_config}_i = \operatorname{round}(\max(x_i \cdot N,\ \textit{min\_sats}))
\]
where $\operatorname{round}(k)$ returns the integer nearest to $k$ for any $k \in \mathbb{R}$. Here, $\text{min\_sats}$ is an optional minimum number of satellites required per orbit. If the integer allocations do not sum to $N$ due to rounding, we penalize the resulting orbit (mapping to a negative entanglement generation rate), ensuring that the optimization process naturally avoids infeasible solutions without losing the invertibility afforded to us by the ALR transformation.

\subsection{Complete parameter space}
\label{sec:p_space}

The result is a parameter space that includes both inclination angles and satellite allocations:
\begin{itemize}
    \item Inclinations $\theta_i \in [0^\circ, 180^\circ]$ are directly optimized as continuous variables.
    \item Satellite allocations are represented via unconstrained real parameters $(y_1,\ldots,y_{D-1})$ through the ALR transformation, ensuring feasibility and smoothness.
\end{itemize}

Our BO and GA optimization approaches each operate on this search space, while respecting the required constraints.

\subsection{Bayesian optimization}

Bayesian optimization is particularly suitable for problems where the objective function is difficult to compute\cite{frazier2018tutorialbayesianoptimization} and the goal is to minimize the number of evaluations while effectively exploring the parameter space. Due to the black box nature and expensive evaluation time, BO provides a promising approach to iteratively refine parameters. Our approach relies on the parameterization described above, objective evaluation, and iterative optimization using a Gaussian process model described in Section \ref{sec:how_bayesian_opt_works}.

\subsection{Genetic optimization}
% \begin{toadd} Look at this? More concise definition/one paragraph\end{toadd}
Genetic algorithms are well-suited for optimization in large non-convex search spaces \cite{doi:10.2514/3.25195}. Our GA framework begins with a diverse population of candidate satellite orbits utilizing the parametrization detailed in Sections~\ref{sec:alr}--\ref{sec:p_space}. Guided by tuned hyperparameters across successive generations, GA offers a viable approach to exploring the search space and uncovering diverse, high-performing configurations.

    \section{Optimization Methodology}\label{sec:opt-methodology}
\begin{figure}
  \centering
  \parbox[c][0.21\textwidth]{0.02\textwidth}{(a)}
  \parbox[c][][b]{0.21\textwidth}{\includegraphics[width=0.21\textwidth]{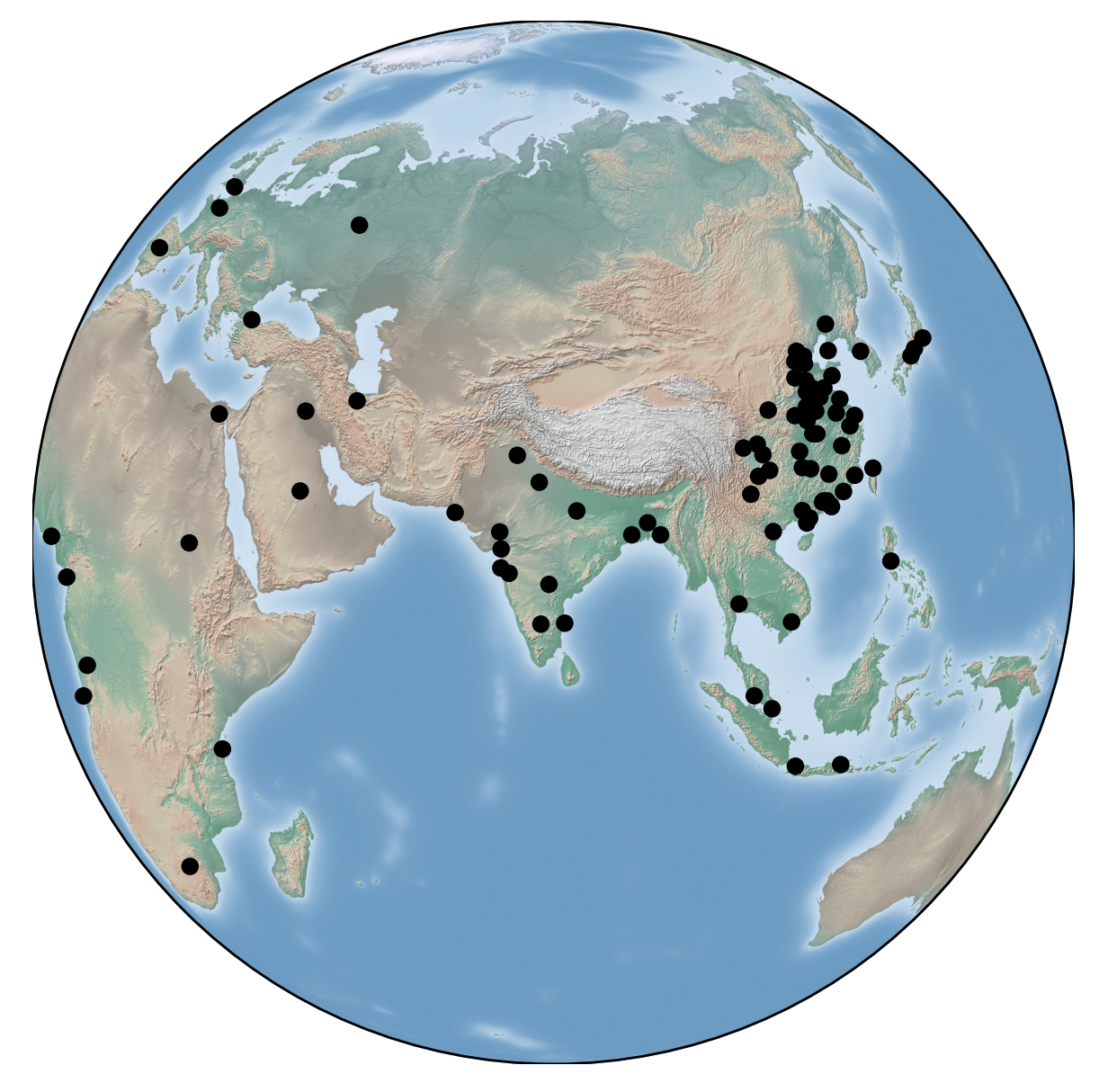}}
  \parbox[c][0.21\textwidth]{0.02\textwidth}{(b)}
  \parbox[c][][b]{0.21\textwidth}{\centering\includegraphics[width=0.21\textwidth]{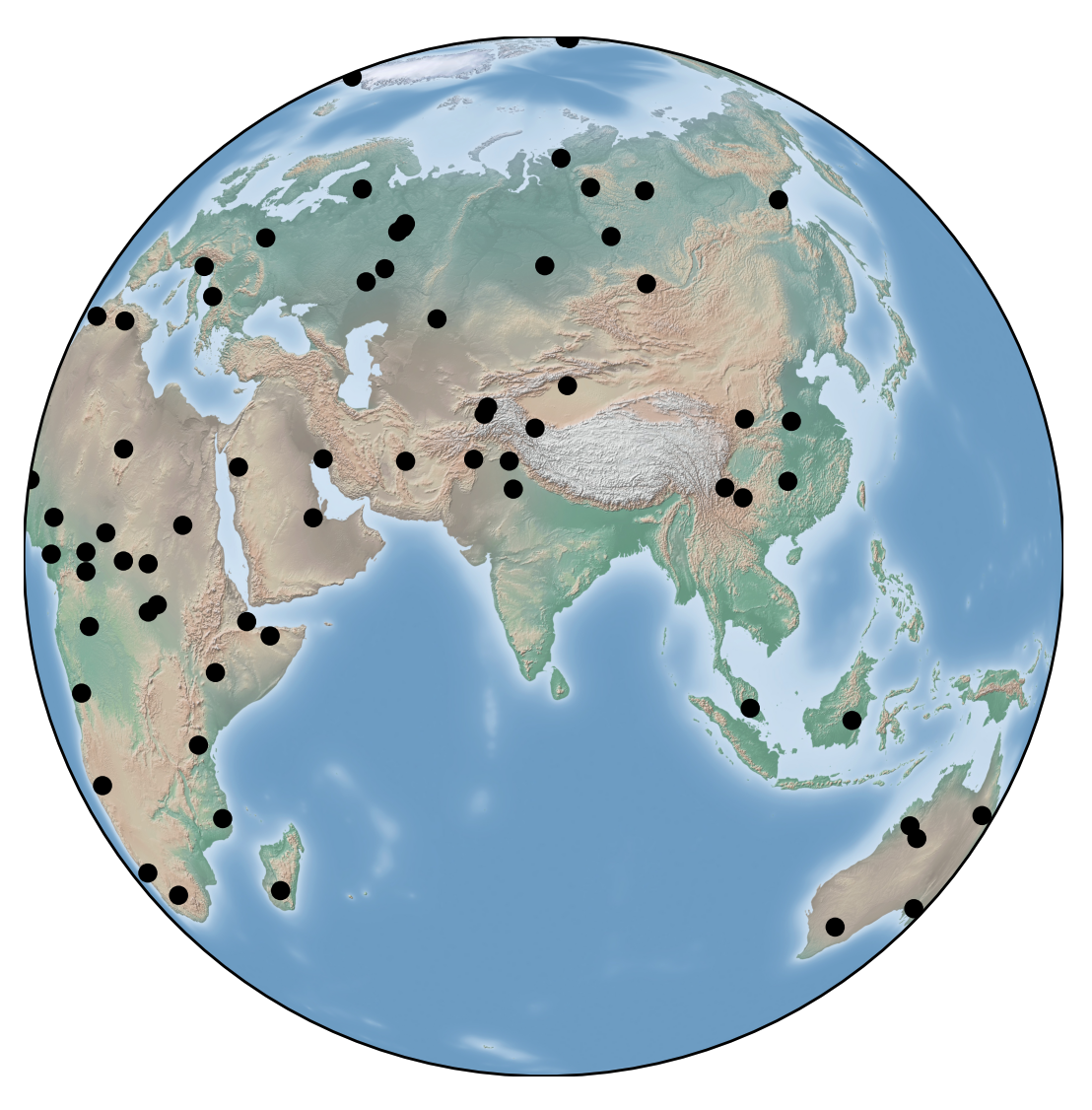}}

  \captionsetup{width=0.49\textwidth}%
  \caption{%
    (a) 100 ground stations placed at high-density population centers
    (b) 100 ground stations placed randomly with a fixed seed exclusively on land
  }
  \label{fig:sat-gs-info}
\end{figure}
\subsection{Bayesian optimization}\label{sec:how_bayesian_opt_works}
\begin{figure}
    \centering
    \includegraphics[width=1\linewidth]{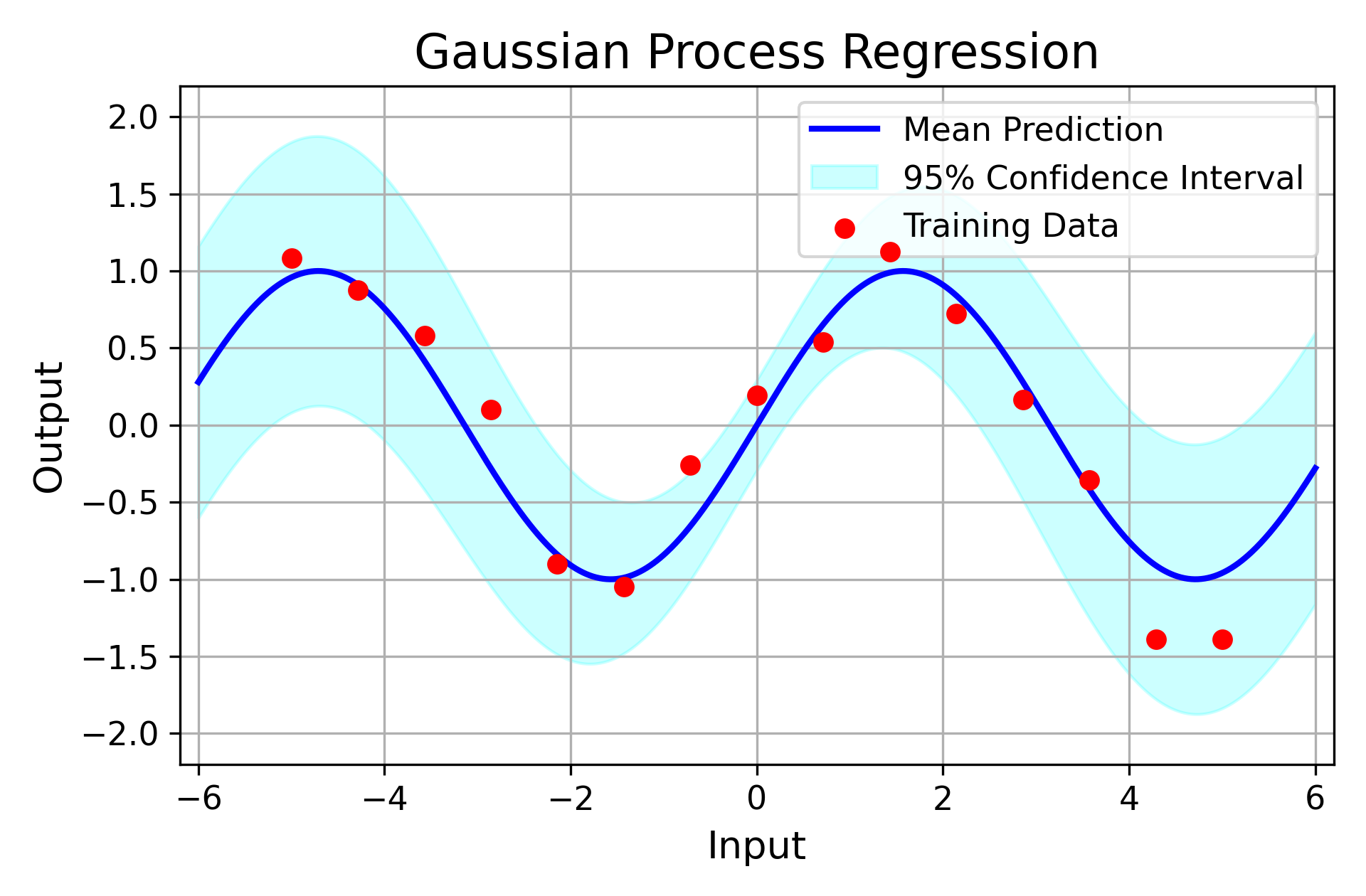}
    \caption{An example of a Gaussian Process}
    \label{fig:GP}
\end{figure}
\begin{algorithm}
\caption{Bayesian Optimization Framework}
\label{alg:bo}
\begin{algorithmic}[1]
\Procedure{$\mathtt{BO}$}{$\mathbf{x}, \text{rate}$}

\# \textbf{$\mathbf{x} = (\theta_1, \dots, \theta_D, y_1, \dots, y_{D-1})$}

    \State \textbf{Add $(\mathbf{x}, \text{rate})$ to dataset $\mathcal{D}$}
    
    \quad \# Store latest evaluated configuration

    \State \textbf{Fit a Gaussian Process model on $\mathcal{D}$}
    
    \quad \# Using Matérn kernel \cite{dowling2021hidamaternkernel}

    \State \textbf{Optimize GP hyperparameters}
    
    \quad \# Maximizes log marginal likelihood 
    
    \quad \# (i.e., L-BFGS-B)

    \State \textbf{Compute posterior mean $\mu(\mathbf{x})$}

    \State \textbf{Compute posterior variance $\sigma^2(\mathbf{x})$}

    \State \textbf{Define (LCB) acquisition function:}
    \State \quad $\alpha(\mathbf{x}) = \mu(\mathbf{x}) - \kappa \cdot \sigma(\mathbf{x})$

    \State \textbf{Select the next configuration to evaluate}
    \State \quad \vspace{0.5em} \textbf{$\mathbf{x}^* \gets \arg\max \alpha(\mathbf{x})$}
    
\# Return next ALR-transformed candidate
    \State \textbf{\Return $\mathbf{x}^*$}
\EndProcedure
\end{algorithmic}
\end{algorithm}
We implement Bayesian optimization \cite{frazier2018tutorialbayesianoptimization} as a probabilistic model to optimize our entanglement distribution rate, which is expensive to calculate. At its core, a given BO implementation uses a Gaussian Process (GP) (see Figure~\ref{fig:GP}) and an Acquisition Function (AF). After observing simulation results for a set of satellite parameters, the GP \cite{frazier2018tutorialbayesianoptimization} is used as a surrogate model to approximate our objective, providing a predictive distribution. An acquisition function \(\alpha(\mathbf{x})\) then proposes its next ideal constellation. We evaluate its constrained form, repeating this process in full until we reach our defined number of iterations.

For our satellite optimization task, we first sample $\textbf{x} = (\theta_1, \dots, \theta_D, y_1, \dots, y_{D-1})$  at random for the first $n$ iterations. Then, given the AF-proposed constellation and corresponding rate, we fit a Gaussian Process model to our dataset using a Matérn kernel \cite{dowling2021hidamaternkernel}. See algorithm~\ref{alg:bo} for pseudocode detailing this process. 

\begin{figure*}[htbp]
  \centering
  \parbox[t]{0.02\textwidth}{(a)}
  \parbox[t][][b]{0.43\textwidth}{\centering\includegraphics[width=0.43\textwidth]{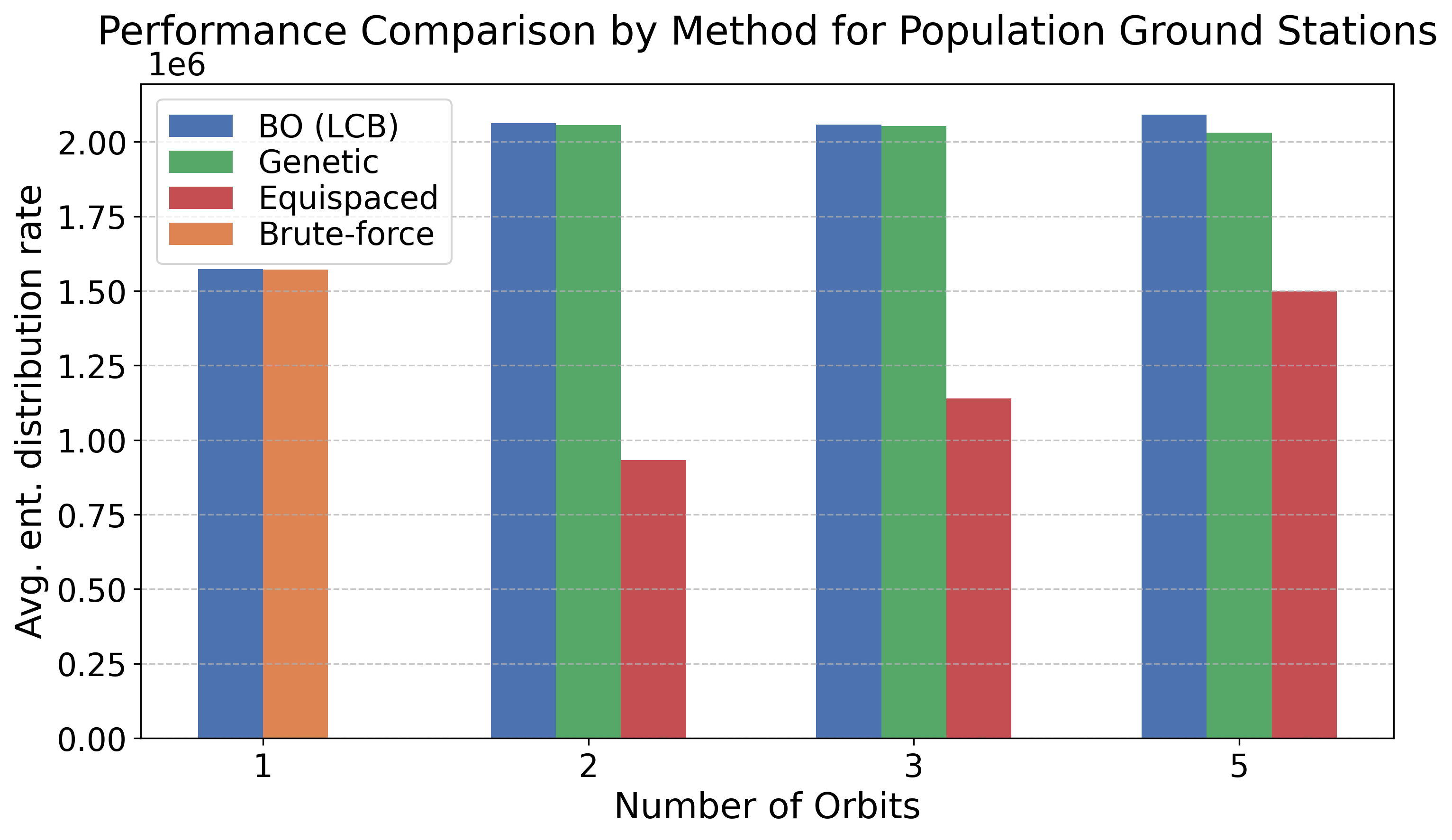}}
  \parbox[t]{0.02\textwidth}{(b)}
  \parbox[t][][b]{0.43\textwidth}{\centering\includegraphics[width=0.43\textwidth]{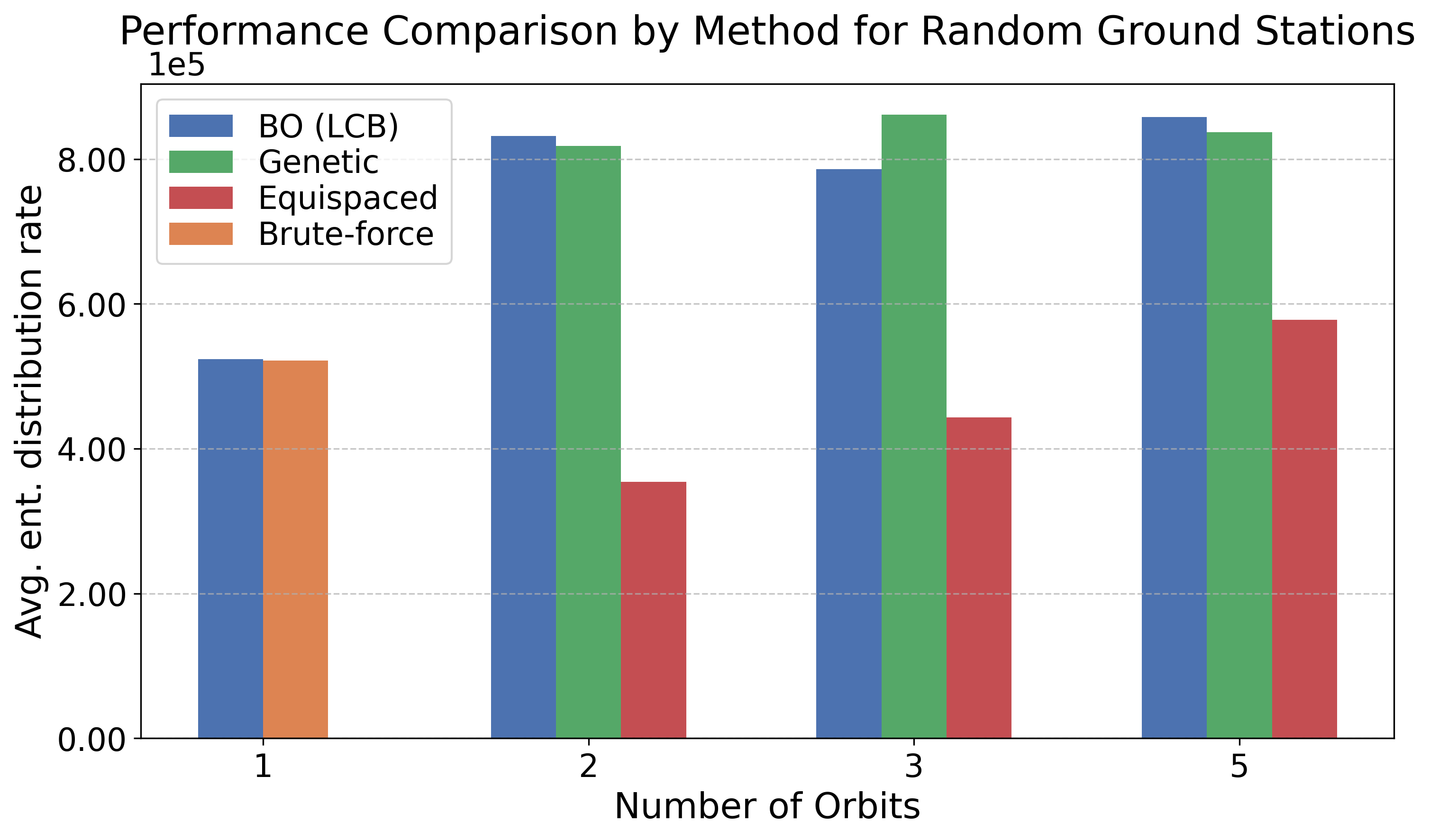}}

  %\vspace{-1em}

  \parbox[t]{0.02\textwidth}{(c)}
  \parbox[t][][b]{0.43\textwidth}{\centering\includegraphics[width=0.43\textwidth]{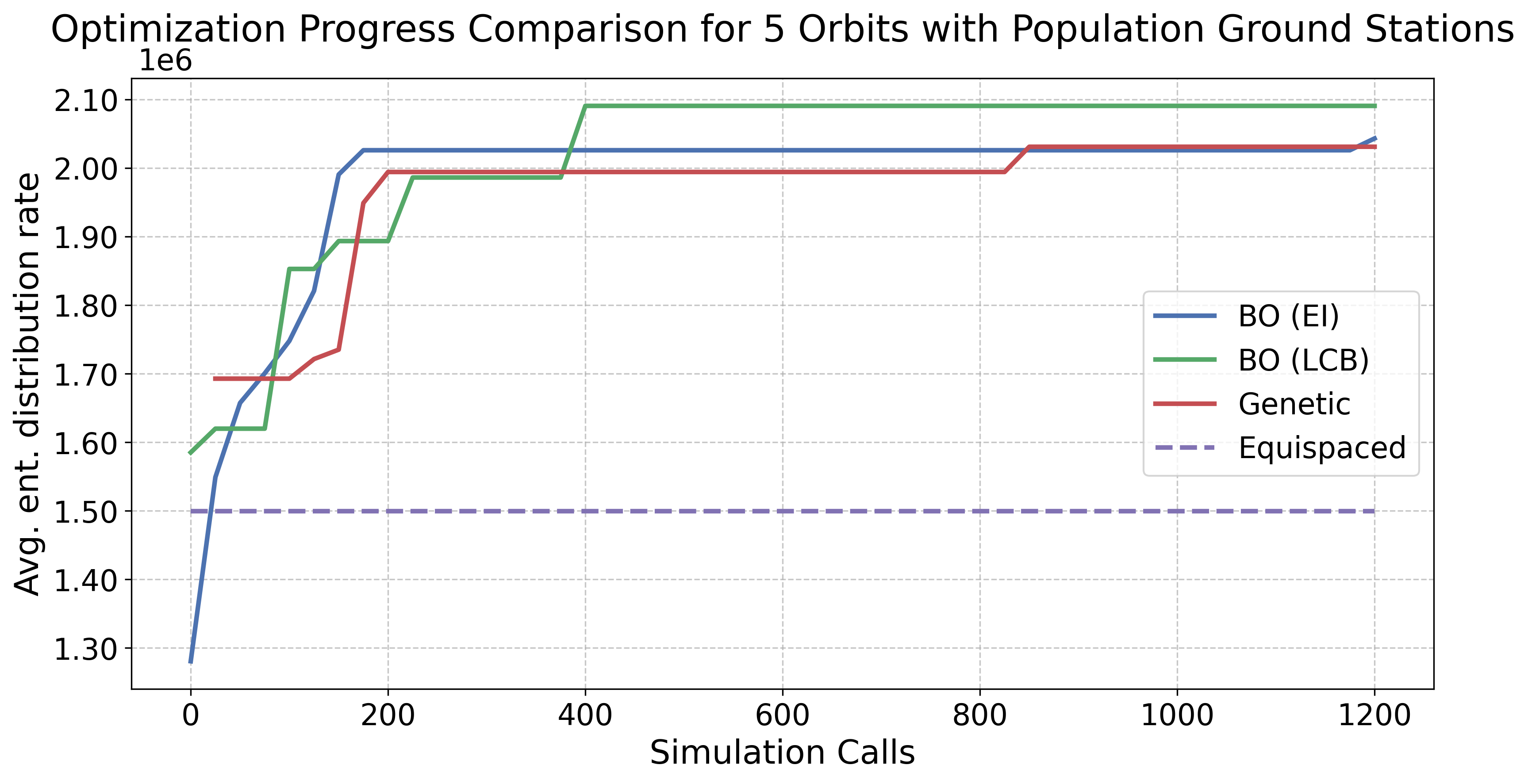}}
  \parbox[t]{0.02\textwidth}{(d)}
  \parbox[t][][b]{0.43\textwidth}{\centering\includegraphics[width=0.43\textwidth]{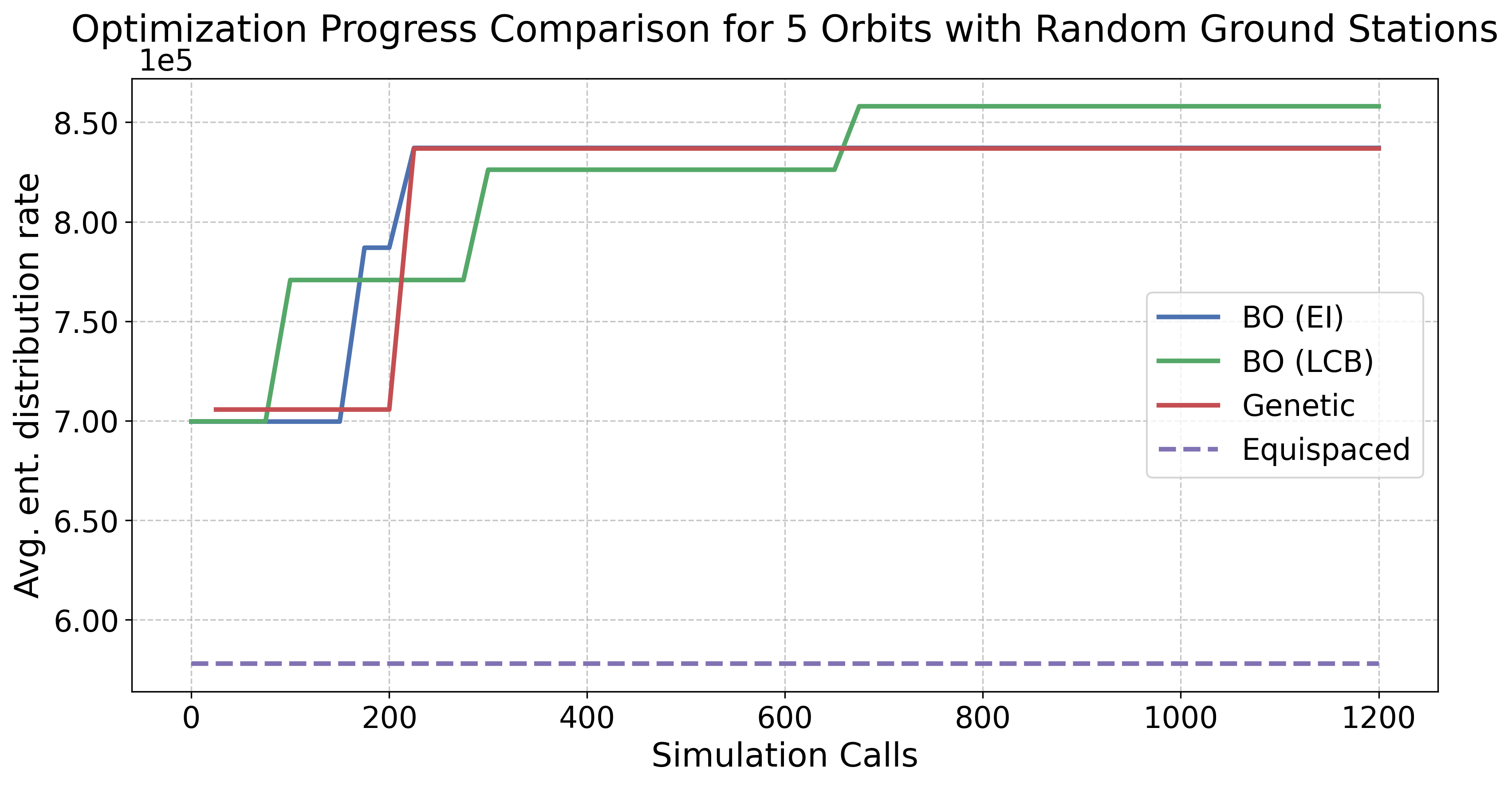}}

  %\vspace{-1em}

  \parbox[t]{0.02\textwidth}{(e)}
  \parbox[t][][b]{0.43\textwidth}{\centering\includegraphics[width=0.43\textwidth]{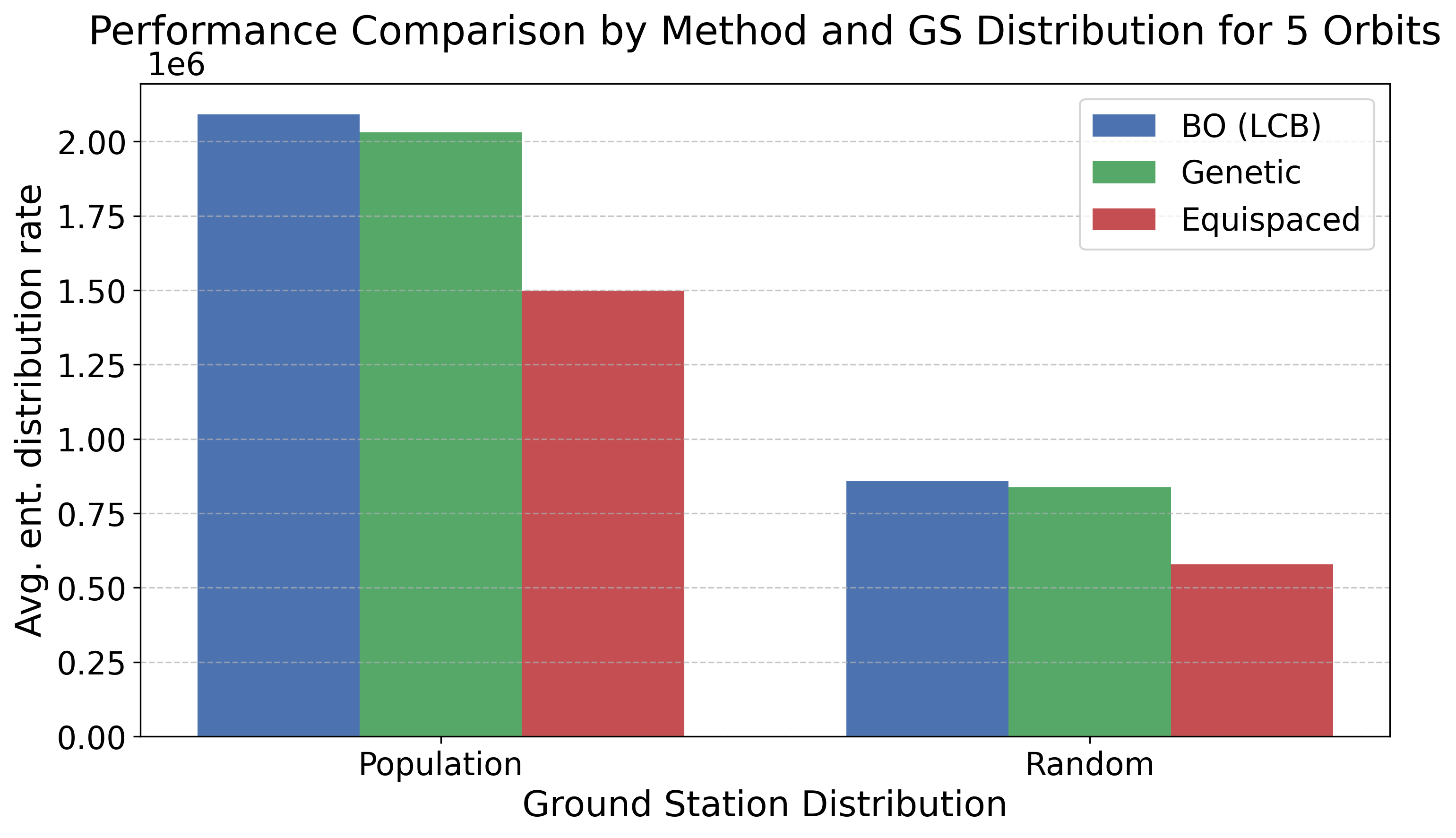}}
  \parbox[t]{0.02\textwidth}{(f)}
  \parbox[t][][b]{0.43\textwidth}{\centering\includegraphics[width=0.43\textwidth]{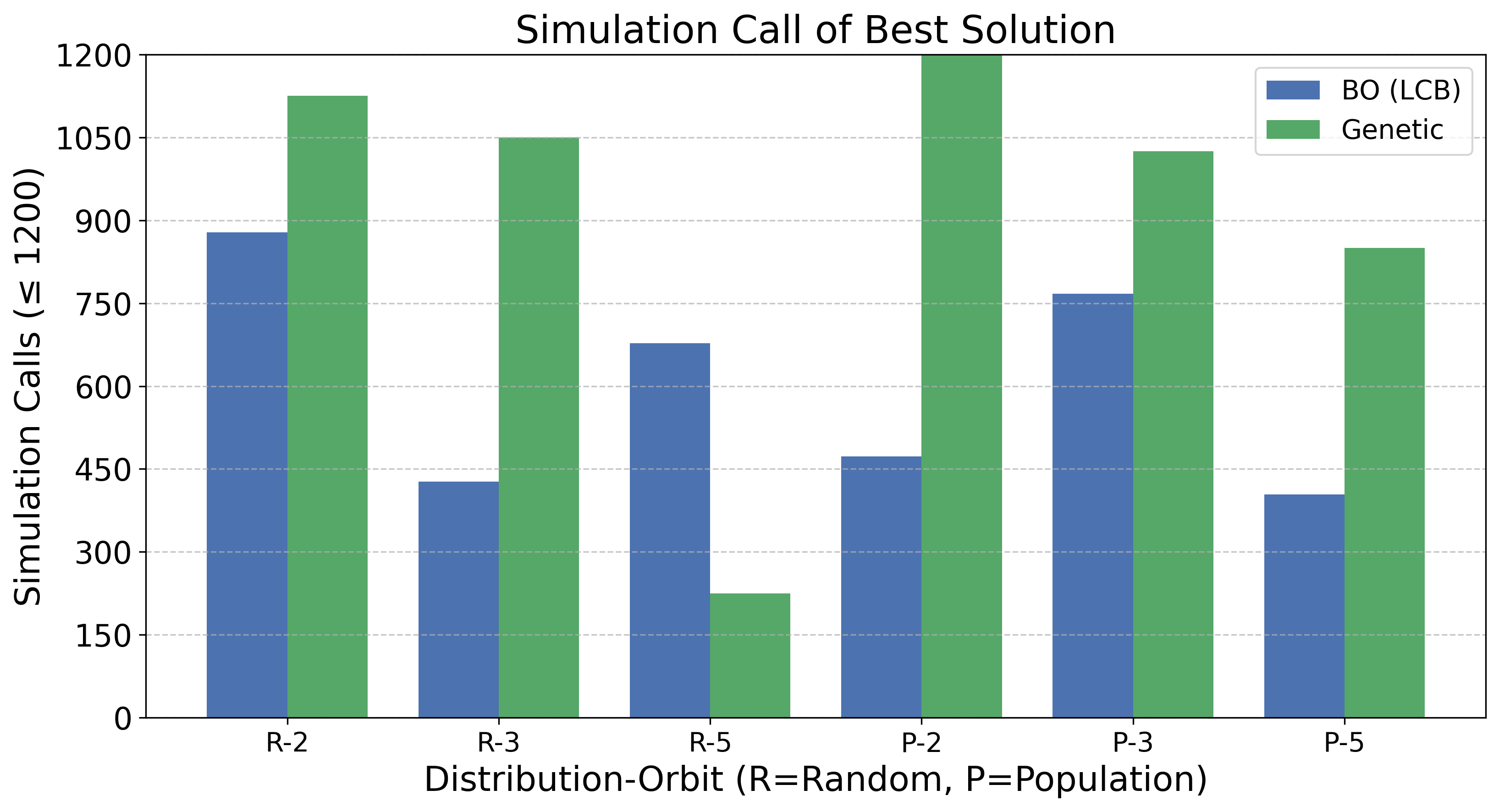}}

  \captionsetup{width=\textwidth}%
  \caption{Performance comparison of optimization methods and baselines across different orbit counts and ground station distributions. 
  (a)–(b): Average entanglement distribution rate (EPR pairs/second) across 1, 2, 3, and 5 orbits using population-based (a) and random (b) land-based ground station placements. The single orbit case benchmarks against a brute-force baseline, while multi-orbit results are compared against equispaced inclination angles.
  (c)–(d): Optimization progress over simulation calls for population-based (c) and random (d) ground stations, showing the maximum rate by the time of a call.
  (e): Comparative performance across methods (BO, GA, equispaced) under different ground station distributions for 5 orbits. (f): Simulation call at which each optimization method first achieved its final best solution, compared across orbit counts (2, 3, and 5) and ground station distribution types (random and population-based).
  }
  \label{fig:six_two_per_row}
\end{figure*}

We compare between two of the most common acquisition functions, Lower Confidence Bound (LCB) and Expected Improvement (EI) \cite{9696089}. Due to our search space's high dimensionality and nonuniformity (from regions such that we experience either a rounding error or miss all ground stations) we expect (see \cite{9696089}) and observe LCB to behave more reliably than EI (see Figure~\ref{fig:six_two_per_row}(c), which shows the quality of the best constellation found by each optimization technique up to a given number of simulation calls for population ground stations).

\subsection{Genetic algorithm}
We implement a traditional genetic algorithm that involves fitness evaluation, selection, crossover, and mutation. The population comprises individuals that encode satellite proportion parameters and orbital inclination angles in tuples, initialized and constrained to the parametrization detailed in Section \ref{sec:parametrization}. Fitness is defined on the basis of maximizing an individual's EPR generation rate.

To minimize premature convergence, we incorporate three primary mechanisms. First, the mutation rate follows an exponential decay schedule as a function of generation index \cite{10.5555/93126}. Second, we maintain a parent pool of the top five individuals and randomly sample from this pool during crossover \cite{10.1007/3-540-58484-6_252}. Third, 10\% of each generation is filled with randomly generated individuals to enforce population diversity.

Within each generation, the population update proceeds as follows. Offspring are generated through repeated parent sampling, crossover, and mutation until 90\% of the next population is filled. The remaining 10\% consists of newly sampled random individuals. All individuals are evaluated and ranked. The ordered list of best-performing individuals is retained and propagates through the next generations until termination. See Algorithm~\ref{alg:ga} for pseudocode detailing this process. 

\begin{algorithm}
\# Initialize population and store to dataset $|\mathcal{D}| = 25$\\
\# ALR transformation applied within init\_pop() method \\
\# Track gen\_num to decay mutation rate
\caption{Genetic Algorithm Framework}
\label{alg:ga}
\begin{algorithmic}[1]
\Procedure{$\mathtt{GA}$}{$\mathbf{x}, \text{rate}$}

    \# Procedure only uses $\mathcal{D}$ and gen\_num.

    \# \textbf{$\mathbf{x} = (\theta_1, \dots, \theta_D, y_1, \dots, y_{D-1})$}

\For{$\mathbf{x_i} \gets \mathcal{D}$}
    \State $\text{fitnesses[i]} \gets \mathtt{evalute\_individual}(\mathbf{x_i})$
\EndFor
\State $\text{parents} \gets \mathtt{select\_parents}(\text{fitnesses}, \mathcal{D}, 5)$
\State \vspace{0.5em} $\mathtt{update\_mutation\_rate}(\text{gen\_num})$
\While{$|\text{new\_pop}| < 0.90 \times \text{pop\_size}$}
        \State $(\text{par1}, \text{par2}) \gets \mathtt{random.sample}(\text{parents}, 2)$
        \State $\mathbf{x} \gets \mathtt{crossover}(\text{par1}, \text{par2})$
        \State $\mathbf{x} \gets \mathtt{mutate}(\text{child})$
        \State Append $\mathbf{x}$ to $\text{new\_pop}$
    \EndWhile
    \State $\text{num\_random} \gets \text{pop\_size} - |\text{new\_pop}|$
    \State $\text{rand\_individuals} \gets \mathtt{init\_pop}()[:\text{num\_random}]$
    \State \vspace{0.5em} $\text{new\_pop} \gets \text{new\_pop} \cup \text{rand\_individuals}$
    \State $\mathbf{x}^* \gets \mathcal{D}[\mathtt{argmax}(\text{fitnesses})]$
    \State $\mathcal{D} \gets \text{new\_pop}$
    \State \textbf{\Return $\mathbf{x}^*$}
\EndProcedure
\end{algorithmic}
\end{algorithm}

\begin{figure}
  \centering  \includegraphics[width=0.49\textwidth]{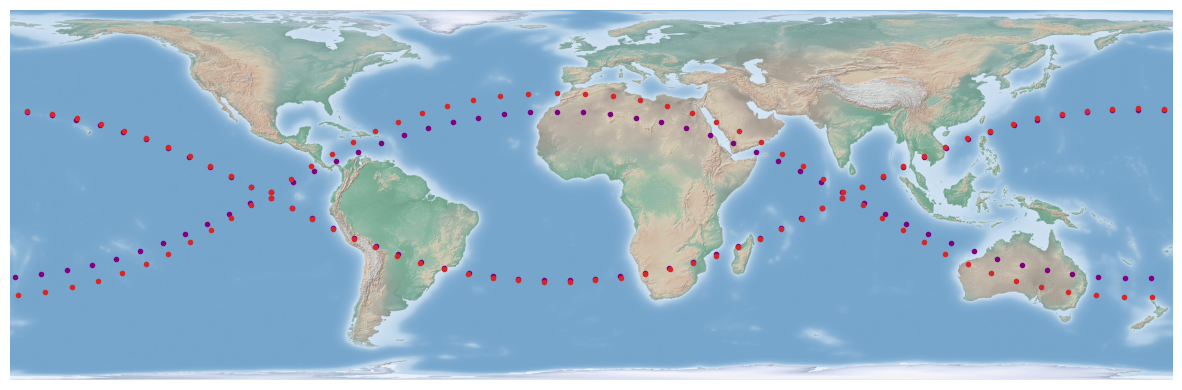}
  \captionsetup{width=0.49\textwidth}%
  \caption{%
    Map visualizations of the optimal satellite configurations yielded by GA (red) and BO (purple) for 2 orbits over population clusters.
  }
  \label{fig:map_bayes_genetic}
\end{figure}

\section{Assessment of Optimization Frameworks}\label{sec:assessment}\vspace{-0.05em}
To assess the effectiveness of the proposed optimization frameworks, we evaluate them for:
\begin{itemize}
    \item 2, 3, and 5 different inclination angles (orbits)
    \item 100 total satellites
    \item 100 ground stations
    \item 1200 simulator calls -- each call comprises 2880 samples collected over a simulated day
\end{itemize}
We evaluate the frameworks under two ground station distributions: one based on population density and one random distribution, as shown in Figures \ref{fig:sat-gs-info}(a) and \ref{fig:sat-gs-info}(b).

We evaluated several naive constellation designs to establish a baseline comparison. Specifically, we tested three trials of 100 satellites at equispaced angles over $0^\circ$ to $180^\circ$ across 2, 3, and 5 distinct inclinations. These reflect the parameter settings we trained for comparison and are inspired by Walker constellations \cite{HUANG2021151}, which have shown to provide the foundation for achieving continuous global coverage with a minimal number of satellites.

Last, we conducted a brute-force trial for comparison, testing all 100 satellites in a single orbit and systematically evaluating their performance at 180 different inclination angles, increasing from $0^\circ$ to $179^\circ$ in increments of $1^\circ$.

\section{Numerical Results and Analysis}\label{sec:results}
We evaluate the performance of BO and GA across several optimization trials, focusing on their respective entanglement generation rates (EPR pairs/second).
\subsection{Effect of orbit count}
Increasing the number of orbits enhances performance. For population-based ground station layouts:
\begin{itemize}
\item The best single orbit configuration achieves approximately $1.57 \times 10^6$ EPR/s
\item Transitioning to 2 orbits yields a 30$\%$ increase, surpassing $2.06 \times 10^6$ EPR/s
\item 3-orbit and 5-orbit configurations deliver at best marginally higher rates (up to $2.09 \times 10^6$), though returns diminish. 

\end{itemize}
Notably, over random ground stations, 5-orbit GA fails to replicate the strong clustering observed in 3-orbit solutions, and 3-orbit BO fails to even reach the optimal 2-orbit split, prematurely converging. We hypothesize two well-distributed constellations are sufficient to realize most entanglement benefits, with further orbits offering limited improvements given the constellation size.

\subsection{Comparison to baseline approaches}
We benchmark our methods against two baselines: a brute-force search for  single orbit configurations and uniformly equispaced inclinations for multi-orbit setups.
\begin{itemize}
\item Brute-force and BO (LCB) both achieve rates of $1.57 \times 10^6$ (population) and $5.22 \times 10^5$ (random) EPR/s for single orbit cases, suggesting that BO is capable of locating global optima in a simple scenario. We excluded GA from the brute-force evaluation because a single orbit prevents it from meaningfully applying crossover, causing the algorithm to reduce to a randomized hill-climbing search.
\item Equispaced inclinations across 5 orbits yield $1.50 \times 10^6$ EPR/s for population layouts --- well below the $2.09 \times 10^6$ EPR/s achieved by optimized methods.
\end{itemize}
These comparisons, as visualized in the histograms of network EPR rate by number of orbits and method in Figures \ref{fig:six_two_per_row}(a) and \ref{fig:six_two_per_row}(b), highlight the clear advantage of ground-station aware optimization over naive strategies, particularly in complex multi-orbit setups.

\subsection{Bayesian optimization vs. genetic algorithm}

On an M1 Pro 10-core CPU, a simulated day for an equispaced orbit runs in roughly 180 seconds. As a general estimate, we can then expect 1200 simulation calls to take around 60 hours. Figures \ref{fig:six_two_per_row}(c) and \ref{fig:six_two_per_row}(d) illustrate the timeline of convergence for each method by showing the entanglement rate of the best constellation found by each optimization technique in up to a given number of simulation calls for population ground stations and random ground stations respectively. We analyze convergence behavior and performance of BO and GA across simulation calls:
\begin{itemize}
\item BO rapidly approaches optimal configurations: $2.09 \times 10^6$ EPR/s (population) by simulation call 404 and $8.26 \times 10^5$ (random) by iteration 301. For the latter, by iteration 678, it exceeds $8.58 \times 10^5$, but there is no improvement with additional runs.
\item GA similarly converges to nearly the same value, requiring approximately 200 simulation calls for the 5-orbit random case, but occasionally surpassing BO as runtimes increase. For instance, its 3-orbit result of $8.60 \times 10^5$ is reached at simulator call 1050, while BO's maximum remains at $7.86 \times 10^5$ from as early as call 427. Similar behavior can be seen for population ground stations in Figure \ref{fig:six_two_per_row}(c), which shows an increase beyond 800 iterations.
\end{itemize}

Figure~\ref{fig:six_two_per_row}(e) presents a bar chart that compares the best entanglement distribution rate found by each method for each ground station distribution for 5 orbit constellations. For both ground station layouts, both BO and GA significantly outperform the equispaced baseline, while achieving nearly identical optimal entanglement distribution rates. This highlights the importance and reliability of ground-station aware optimization when working with multi-orbit constellation designs.

Figure~\ref{fig:six_two_per_row}(f) shows the simulation call at which the best performing constellation was found for each method as a function of number of orbits and ground station distribution. It illustrates BO's tendency to identify strong candidates early. This is markedly demonstrated by all but the 5-orbit random case --- due to the especially early convergence of GA, we speculate that this case is simply an outlier. We note that in our experience, BO always converges within 1000 simulation calls. As such, we hypothesize that BO’s exploration-driven sampling often identifies strong candidates early, though its reliance on Gaussian process assumptions may contribute to premature convergence. GA, by contrast, continues improving gradually, though it sees few substantial gains over this period.

\subsection{Impact of ground station layout}
Ground station placement significantly affects performance.
\begin{itemize}
\item Constellations over population-based ground stations consistently outperform random ones
\item The best result over population ground stations exceeds $2.09 \times 10^6$ EPR/s, while random layouts peak at $8.58 \times 10^5$, reflecting a 2.52$\times$ disparity
\end{itemize}
We attribute this to the clustered nature of population centers, which allows more frequent dual-downlink opportunities and reduced idle satellite coverage.

\subsection{Constellation shape characteristics}
Figure~\ref{fig:map_bayes_genetic} compares the final optimized orbital configurations for BO and GA.
\begin{itemize}
    \item Both methods converge to structurally similar constellations, typically involving two dominant orbits and one or two auxiliary ones
    \item BO's best configuration over population based ground stations uses 45, 37, 15, 1, and 2 satellites with inclinations of 153.21, 24.72, 143.04, 69.94, and 149.05 degrees
    \item GA's top configuration over population based ground stations consists of 54, 33, and 13 satellites with inclinations of 150.53, 24.39, and 40.32 degrees
\end{itemize}
The consistency across methods suggests the existence of stable optima in the configuration space -- which both methods can reliably approach, given sufficient time.

\subsection{Summary}
We conclude that both BO and GA are highly effective tools for this optimization task. We further speculate, having observed that BO rarely improves after 800 iteration calls, that BO is highly effective for rapid optimization, especially when iteration budgets are limited, but is more susceptible to local optima. GA, while slower to converge (often requiring over 1000 iterations to find maximum values), continues to improve and can match or exceed BO when allowed to run longer. Population-based station placements dramatically outperform random layouts, and most performance benefits arise with just two well-chosen orbits for constellations of this size. 

\section{Conclusion \& Future Directions}\label{sec:conclusion}
In this work, we explored how existing black-box optimization frameworks can be adapted to optimize satellite placement in constellations for entanglement distribution. Specifically, we use bayesian optimization and genetic algorithms to tune orbital inclinations and satellite cluster allocations to distribute entanglements to a known set of ground stations. By carefully evaluating these approaches through extensive simulations on two different ground station placements and a multitude of satellite configuration settings, we show that these methods can be highly efficient for searching this high-dimensional design space. From our findings, we observe that careful selection of orbital parameters can substantially improve entanglement distribution rates when informed by the ground station infrastructure. We demonstrate that optimizing even a small number of inclination angles, while balancing satellites across those orbits, is an effective way to boost the overall performance of the constellation.

 A natural next step involves generalizing the optimization approaches examined here to diverse operational contexts and mission-specific goals, such as maximizing quantum key distribution rates or some metrics of fairness rather than solely maximizing entanglement distribution rates. We also would like to explore the constellation design problem with more degrees of freedom, such as allowing variations in satellite altitude, including unmanned aerial vehicle and high altitude platform entanglement sources and relays, or simulating satellites with onboard quantum memories for inter-satellite links.

\bibliography{bib}

\end{document}